\documentclass[12pt]{iopart}
\usepackage{graphicx}
\begin{document}

\title[Spontaneous creation of chromomagnetic field and $A_0$-condensate on a lattice]
{Spontaneous creation of chromomagnetic field and $A_0$-condensate at high temperature on a lattice}

\author{V.Demchik, V.Skalozub}

\address{Dnipropetrovsk National University, Dnipropetrovsk, Ukraine}
\ead{vadimdi@yahoo.com, skalozub@ff.dsu.dp.ua}

\begin{abstract}
In a lattice formulation of  $SU(2)$-gluodynamics, the spontaneous
generation of  chromomagnetic fields at high temperature is
investigated. A procedure to determine  this phenomenon is
developed.  By means of the $\chi^2$-analysis of the data set
accumulating $5-10\times 10^6$ Monte Carlo configurations, the
spontaneous creation of the Abelian color magnetic field is
indicated. The common generation of the magnetic field and
$A_0$-condensate is also studied. It is  discovered that the field
configuration consisting of the magnetized vacuum and the
$A_0$-condensate is stable.
\end{abstract}

\vspace{2pc}

\section{Introduction}
Nowadays it is generally accepted that the non-linearity of
non-Abelian gauge fields could result in formation of  field
condensates. The first model of gluon condensate, so called {\it
"color ferromagnetic vacuum state"}, was proposed thirty years ago
by Savvidy \cite{SG}.  It describes the spontaneous generation of
the uniform Abelian chromomagnetic field $H=const$ due to a vacuum
polarization. Unfortunately, this state is unstable because of the
tachyonic mode in the gluon spectrum \cite{Skalozub78, Olesen78}.
This situation is changed at finite temperature when the  spectrum
stabilization  happens due to either a gluon magnetic mass
\cite{SB} or a so-called $A_0$-condensate \cite{SZ} that is
implemented in a stable magnetized vacuum. These are the
extensions of the Savvidy model to the finite temperature case
investigated already by the methods of continuum field theory
\cite{EO,SZ,SB}. In these ways the possibility of the spontaneous
generation of strong temperature-dependent and stable color
magnetic fields of order $gH \sim g^4 T^2$, where $g$ is a gauge
coupling constant, $T$ - temperature,  is realized. The field
stabilization is ensured by the temperature and field dependent
gluon magnetic mass, which serves as a regulator of infrared
singularities at $T\not = 0$.

Such spontaneously created "primordial" color magnetic fields had
been gene\-rated at a GUT scale \cite{PL}. They could serve as
seed fields responsible for generation of the large-scale magnetic
fields detected in  astrophysical observations \cite{GR}. The
presence of strong magnetic fields in the early universe is of
paramount importance for its evolution. In particular, a present
day cosmic magnetic field of order $B_0\sim (4-5)\times 10^{-9} G$
could produce the recently discovered Wilkinson Microscopic
Anisotropy Probe (WMAP) anomaly \cite{Campanelli}.

In Refs. \cite{SB,S2} the spontaneous creation of the
chromomagnetic fields  was observed in $SU(2)$- and
$SU(3)$-gluodynamics within the one-loop plus daisy resummations.
In Ref. \cite{AG} the chromomagnetic condensate of the same order
was calculated in a stochastic QCD vacuum model  by comparison
with some  data of lattice simulations. In Ref. \cite{DSlat} the
spontaneous generation of the chromomagnetic field at high
temperature was investigated in a lattice formulation of
$SU(2)$-gluodynamics. In Refs. \cite{CC} the response of the
vacuum to the influence of strong external fields at different
temperatures was investigated and it has been shown that
confinement is restored when  the strength of the external field
is increased.

Other condensate, which  formation at $T\not =0$ was investigated
in continuum field theory  \cite{RevA0}, is  the zero component of
the gluon electrostatic potential $A_0=const$. It acts to
stabilize the QCD vacuum magnetized state, as it was discussed in
Ref. \cite{SZ}. In this paper, however, the magnetic field was
considered as external one.  Whether or not the actual value of
the $A_0$  generated in the deconfining phase is sufficient to
remove the tachyonic instability  was not estimated.

The main goal of present paper is to investigate  the common
generation of the chromomagnetic field and $A_0$-condensate in
lattice simulations. To incorporate the chromomagnetic field on a
lattice we  use the method developed in \cite{DSlat}. Instead of
the field strength, which is quantized, the magnetic fluxes are
considered as the  objects to be investigated. The flux takes
continuous values. Therefore the minimization of free energy of
the flux  can be done in a usual known procedure. The values of
the strength of spontaneously generated magnetic field and the
values of the Polyakov loop for corresponding states were obtained
from  Monte Carlo (MC) simulations. Then the value of $A_0$ was
derived by using a procedure developed in Ref.\cite{Meis}.

The paper is organized as follows. In section 2, necessary
information about the chromo\-magnetic fluxes on a lattice is
educed and  the results of calculations are given.  In section 3,
the investigation of the effect of combined generation of
chromomagnetic field and $A_0$-condensate  is presented. The final
section is devoted to discussion.

\section{Chromomagnetic fields on a lattice}
In continuum, to determine the spontaneous generation of a
magnetic field one has to minimize the effective potential (or
free energy) of the  background field. The background field is
introduced by splitting of the gauge field potential $A_\mu$ into
the quantum $A_\mu^R$ and classical $\bar{A_\mu}$ parts:
$A_\mu=A_\mu^R+\bar{A_\mu}$. We choice the   potential
$\bar{A_\mu}^a=(0,0,Hx^1,0)\delta^{a 3}$ that corresponds to a
constant chromomagnetic field directed along the third axis in the
Euclidean and color space.

We relate the free energy density of the flux to the effective
action according to the definition,
\begin{eqnarray}
\label{enactions} F(\varphi)=\bar{S}(\varphi)-\bar{S}(0),
\end{eqnarray}
where $\bar{S}(\varphi)$ and $\bar{S}(0)$ are the effective
lattice actions with and without chromomagnetic field,
correspondingly, $\varphi$ is the field flux.

To detect the spontaneous creation of the field it is necessary to
show that  free energy  has a global minimum at  non-zero flux,
$\varphi_{min}\not =0$.

In what follows, we use the hypercubic lattice $L_t\times L_s^3$
($L_t<L_s$) with the hypertorus geometry; $L_t$ and $L_s$ are the
temporal and the spatial sizes of the lattice, respectively. In
the limit of $L_s \to \infty$ the temporal size $L_t$ is related
to physical temperature. The standard Wilson action of the $SU(2)$
lattice gauge theory can be written as
\begin{eqnarray}
\label{Wilson} S_W=\beta\sum_x\sum_{\mu>\nu}\left[1-\frac12 \Tr
U_{\mu\nu}(x)\right];\\\label{plaq}
U_{\mu\nu}(x)=U_\mu(x)U_\nu(x+a\hat{\mu})U_\mu^\dagger(x+a\hat{\nu})U_\nu^\dagger(x),
\end{eqnarray}
where $\beta=4/g^2$ is the lattice coupling constant, $g$ is the
bare coupling, $U_\mu(x)$ is the link variable located on the link
leaving the lattice site $x$ in the $\mu$ direction,
$U_{\mu\nu}(x)$ is the ordered product of the link variables. The
effective action $\bar{S}$ in (\ref{enactions}) is the Wilson
action $S_W$ averaged over the Boltzmann configurations produced
in the MC simulations.

The lattice variable $U_\mu(x)$ can be decomposed in terms of the
unity, $I$, and Pauli, $\sigma_j,$ matrices in the color space,
\begin{eqnarray}\label{compu}
U_\mu(x)=IU_\mu^0(x)+i\sigma_jU_\mu^j(x).
\end{eqnarray}

\begin{figure}
\begin{center}
\includegraphics[bb=58 526 324 779,width=0.35\textwidth]{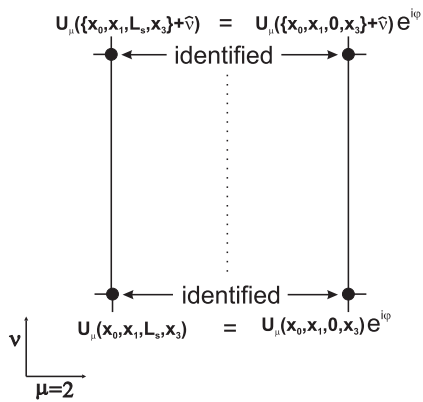}
\end{center}
\caption{The plaquette presentation of the twisted boundary
conditions.}
\end{figure}

The chromomagnetic flux $\varphi$ through the whole lattice was
introduced by applying the twisted boundary conditions \cite{TH}
(Fig. 1). In this approach the edge links in all directions are
identified as usual periodic boundary conditions except for the
links in the second spatial direction, for which the additional
phase $\varphi$ is added. The magnetic flux $\varphi$ is measured
in angular units and can take a value from $0$ to $2\pi$.

The twisted boundary conditions (t.b.c.) for the components of
(\ref{compu}) are
\begin{eqnarray}\label{tbc2}
U^0_\mu(x) \leftrightarrow \Biggl\{
\begin{array}{l}
U^0_\mu\cos\varphi-U^3_\mu\sin\varphi~~~~$for$~x=\{x_0,x_1,L_s,x_3\}~$and$~\mu=2,\\
U^0_\mu\hskip 3.8cm~$for other links$,
\end{array}\\\label{tbc3}
U^3_\mu(x)\leftrightarrow \Biggl\{
\begin{array}{l}
U^0_\mu\sin\varphi+U^3_\mu\cos\varphi~~~~$for$~x=\{x_0,x_1,L_s,x_3\}~$and$~\mu=2,\\
U^3_\mu\hskip 3.8cm~$for other links$.
\end{array}
\end{eqnarray}

The relations (\ref{tbc2}) and (\ref{tbc3}) have been implemented
into the kernel of the MC procedure in order to produce the
configurations with the chromomagnetic flux $\varphi$. Thus, the
flux $\varphi$ is taken into account while obtaining a Boltzmann
ensemble at each MC iteration.

The MC simulations are carried out by means of the heat bath
method. The lattices $2\times 8^3$, $2\times 16^3$ and $4\times
8^3$ at $\beta=3.0$, $5.0$ are considered. These values of the
coupling constant correspond to the deconfinement phase and
perturbative regime. To thermalize the system, 200-500 iterations
are fulfilled. At each working iteration, the plaquette value
(\ref{plaq}) is averaged over the whole lattice leading to the
Wilson action (\ref{Wilson}). Then the effective action is
calculated by averaging over the 1000-5000 working iterations. By
setting a set of chromomagnetic fluxes $\varphi$ in the MC
simulations we obtain the corresponding set of values of the
effective action. The value of the condensed chromomagnetic flux
$\varphi_{min}$ is obtained as the result of  minimization of the
free energy density (\ref{enactions}) in the $\varphi$.

The spontaneous generation of chromomagnetic field is the effect
of order $\sim g^4$ \cite{SB}. The results of MC simulations show
the comparably large dispersion. So, the large amount of the MC
data is collected and the standard $\chi^2$-method for the
analysis of data is applied to determine the effect. We consider
the results of the MC simulations as observed ``experimental
data''.

The effective action depends smoothly on the flux $\varphi$ in the
region $\varphi\sim 0$. So, the free energy density can be fitted
by a quadratic function of  $\varphi$,
\begin{eqnarray}\label{parfit}
\label{ffit} F(\varphi)=F_{min}+b(\varphi-\varphi_{min})^2.
\end{eqnarray}

In Eq.(\ref{ffit} ), there are three unknown parameters,
$F_{min}$, $b$ and $\varphi_{min}$. $\varphi_{min}$ denotes the
minimum position of  free energy, whereas the $F_{min}$ and $b$
are the free energy density at the minimum and the curvature of
the free energy function, correspondingly.

The value $\varphi_{min}$ is obtained as a result of minimization
of the $\chi^2$-function
\begin{eqnarray}
\chi^2(F_{min},b,\varphi_{min})=\sum_i\frac{(F_{min}+b(\varphi_i-\varphi_{min})^2-F(\varphi_i))^2}{D(F(\varphi_i))},\\
D(F(\varphi_i))=\sum_{i\in
bin}\frac{(F(\varphi_i)-\hat{F}_{bin})^2}{n_{bin}-1},
\end{eqnarray}
where $\varphi_i$ is the array of the set of fluxes and
$D(F(\varphi_i))$ is the data dispersion, which can be obtained by
collecting the data into  bins (as a function of $\varphi$);
$n_{bin}$  is a number of points and and $\hat{F}_{bin}$  is a
mean value of free energy density in the considered bin. As it was
determined in the data analysis, the dispersion is independent of
 $\varphi$. The deviation of $\varphi_{min}$ from zero
indicates the presence of the spontaneously generated field.

\begin{table}[b]
\caption{The values of the generated fluxes {\small
$\varphi_{min}$} for different lattices.}
\begin{center}
\begin{tabular}{|c|c|c|c|}
  \hline \rule{0pt}{14pt}
  & $2\times 8^3$ & $2\times 16^3$ & $4\times 8^3$ \\\hline
 \rule{0pt}{14pt}
  $\beta=3.0$ & ${0.019^{+0.013}_{-0.012}}$ & $0.0069^{+0.0022}_{-0.0057}$ & $0.005^{+0.005}_{-0.003}$ \\\hline
 \rule{0pt}{14pt}
  $\beta=5.0$ & ${0.020^{+0.011}_{-0.010}}$ &  &  \\\hline
\end{tabular}
\end{center}
\end{table}

The fit results are given in the Table 1. As one can see,
$\varphi_{min}$ demonstrates the $2\sigma$-deviation from zero.

The 95\% C.L. domain of  parameters $F_{min}$ ($b$ for the right
figure) and $\varphi_{min}$ is represented in Fig. 2. The black
cross marks the position of the maximum-likelihood values of
$F_{min}$ ($b$ for the right figure) and $\varphi_{min}$. It can
be seen that the flux is positively  determined. The 95\% C.L.
area becomes more symmetric with the center at the $F_{min}$, $b$
and $\varphi_{min}$ when the statistics is increased. This also
confirms the results of the carried out fitting.
\begin{figure}[t]
\begin{center}
\includegraphics[bb=123 157 520 543,width=0.3\textwidth]{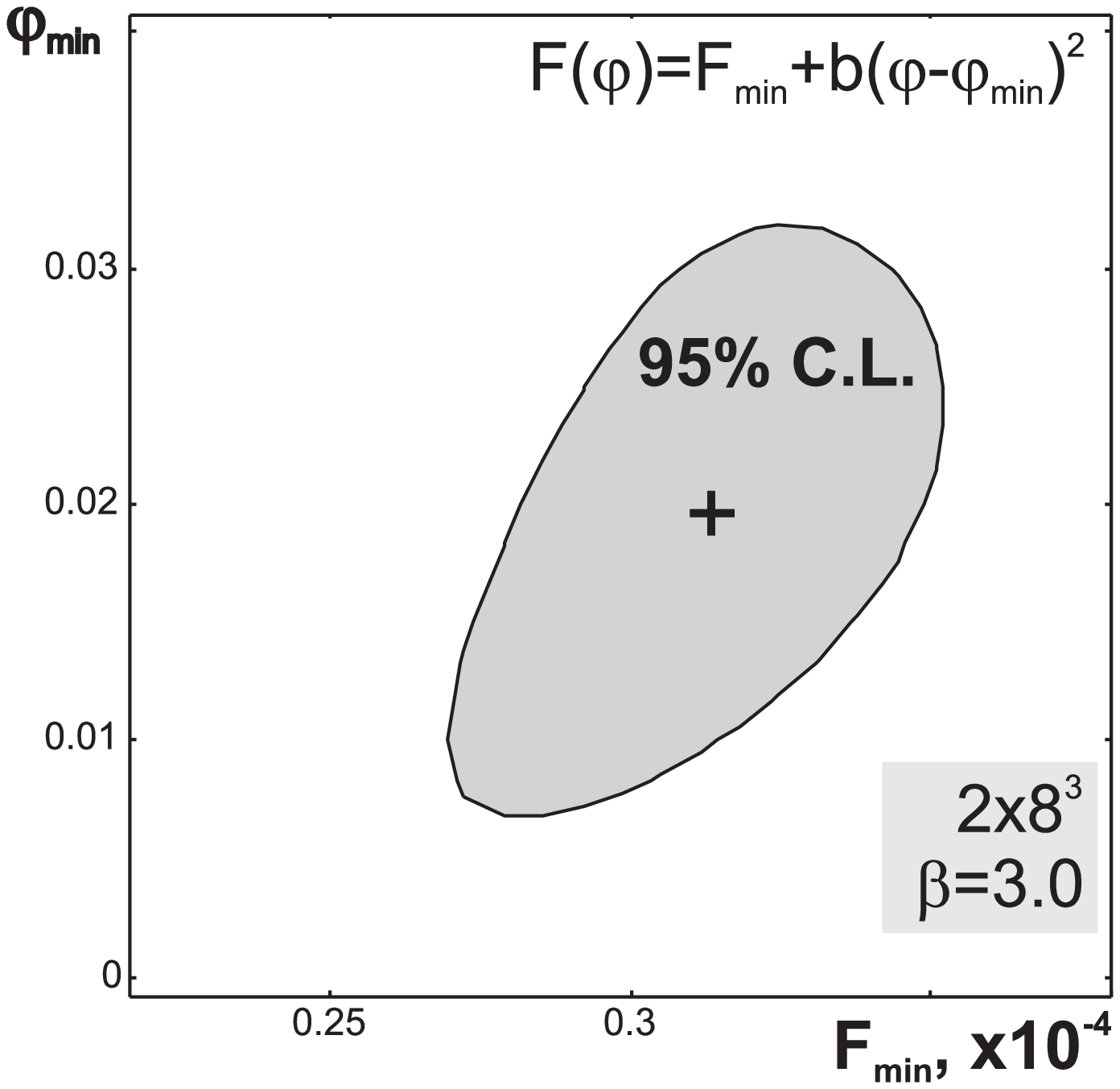}\hskip 2cm
\includegraphics[bb=86 294 471 665,width=0.3\textwidth]{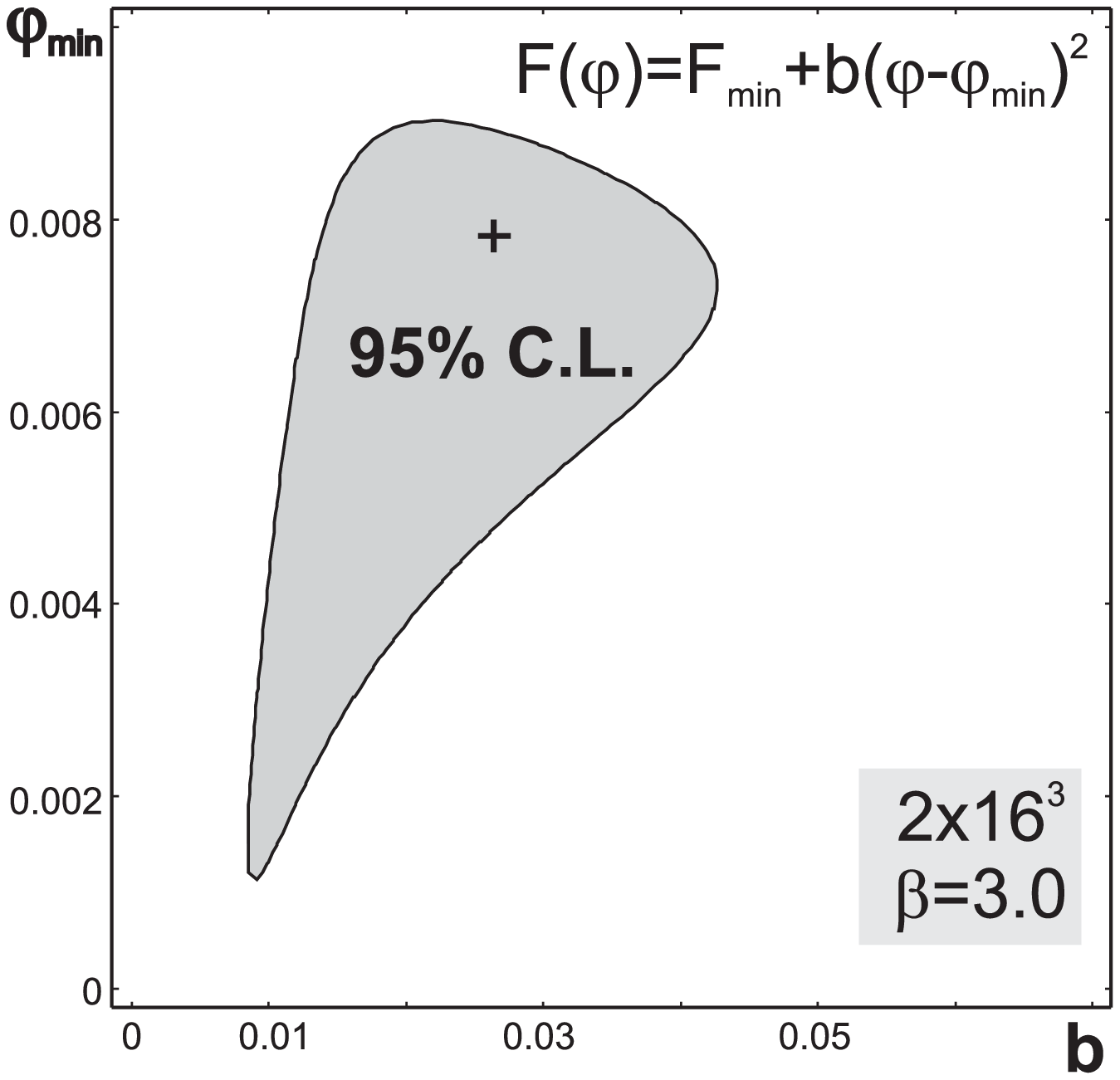}
\end{center}
\caption{The 95\% C.L. area for the parameters $F_{min}$ and
$\varphi_{min}$, determining the free energy density dependence on
the flux $\varphi_{min}$ on lattice $2\times 8^3$ for $\beta=3.0$
(left figure). The 95\% C.L. area for the parameters $F_{min}$ and
$b$ , determining the free energy density dependence on the flux
$\varphi_{min}$ on lattice $2\times 16^3$ for $\beta=3.0$ (right
figure).}
\end{figure}

\section{The spontaneous vacuum magnetization and $A_0$-condensate}
In this section we investigate an effective potential, taking into
consideration  the effects of non-trivial $A_0$-condensate and
 chromomagnetic field \cite{Meis}. This is
realized  by means of calculation of  the partition function in a
general covariant background field gauge with the background field
providing both the  chromomagnetic field and the non-trivial
Polyakov loop. Then the data obtained in  MC simulations are
substituted in this effective potential.  If the minimum value of
it is negative and the potential is real,  the common generation
of  condensates happens.

Within the used imaginary time  formalism, the temporal direction
in the Eucleadian  space is periodic, with period $\beta=1/T$, and
the Polyakov loop $P_L$, defined as the  time-ordered exponential,
\begin{eqnarray}
P_L(\vec{x})=T \exp\left[ig\oint d\tau A_0(\vec{x},\tau)\right],
\end{eqnarray}
is a proper order parameter to describe  the deconfinement phase
transition. It is specified by a constant $A_0$ field, given in
the fundamental representation by
\begin{eqnarray}
A_0=\frac{\phi}{g\beta}\frac{\tau_3}{2},
\end{eqnarray}
where $0\leq\phi\leq 2\pi$, $\tau_3$ is Pauli matrix.

The trace of the Polyakov loop in the fundamental representation
is
\begin{eqnarray}
\Tr_F(P_L)=2\cos(\phi/2).
\end{eqnarray}
Unless the $Z(2) $symmetry is spontaneously broken, the variable
$\phi$  has to have the value $\pi$ corresponding to
$\Tr_F(P_L)=0$.

The one-loop contribution to the free energy has the form
\cite{Meis}
\begin{eqnarray}\label{totEP}
V^{(1)}&=&V^{(1)}_0+V^{(1)}_\pm\\\nonumber
 &=&\sum_n\frac{1}{\beta}\int\frac{d^3\vec{k}}{(2\pi)^3}\log(\omega_n^2
 +\vec{k}^2)+\\\nonumber
 &&\frac12\sum_{m=0}^\infty\sum_{n,\pm}\frac{1}{\beta}
 \frac{gH}{2\pi}\int\frac{dk_3}{2\pi}\log\left[(\omega_n-\frac{\phi}{\beta})^2+2gH(m+\frac12\pm
 1)+k_3^2\right]+\\\nonumber
 &&\frac12\sum_{m=0}^\infty\sum_{n,\pm}\frac{1}{\beta}
 \frac{gH}{2\pi}\int\frac{dk_3}{2\pi}\log\left[(\omega_n+\frac{\phi}{\beta})^2+2gH(m+\frac12\pm
 1)+k_3^2\right],
\end{eqnarray}
where $\omega_n=2\pi n/\beta$ is  the Matsubara frequency,  and
the sum over $n$ is over all integer values. The sum over $m$ is
the sum over the Landau levels.

By applying  a standard technique of Schwinger to express the
logarithms of (\ref{totEP}) as an integral and performing the
high-temperature expansion of the effective potential, we obtain
\begin{eqnarray}\label{totHT}
V^{(1)}=C_1\frac{(gH)^2}{8\pi^2}-\frac{(gH)^{3/2}}{4\pi^{3/2}\beta}
\sum^\infty_{k=2}\frac{2^{2k}B_{2k}}{(2k)!}\Gamma(2k-3/2)\times\\\nonumber
\sum_l\left(\frac{\beta^2 gH}{\beta^2 gH+(\phi-2\pi
l)^2}\right)^{2k-3/2}
+\frac{(gH)^2}{24\pi^2}(3-4\gamma)-\frac{2\pi^2}{45\beta^4}\\\nonumber
+\frac{\phi^2}{3\beta^4}+\frac{\phi^4}{12\pi^2\beta^4}+ \frac{g
H}{2\pi\beta^2}\sqrt{-g H \beta ^2+\phi ^2}+\frac{g^2 H^2}{12 \pi
\sqrt{g H \beta ^2+\phi ^2}}\\\nonumber
-\frac{(gH\beta^2+\phi^2)^{3/2}}{3\pi\beta^2}-\frac{g^2 H^2
\log\left[\frac{\sqrt{-g H} \beta }{4 \pi }\right]}{4 \pi
^2}+\frac{g^2 H^2 \log\left[\frac{\sqrt{g H} \beta }{4 \pi
}\right]}{12 \pi ^2}\\\nonumber -\frac{g^2 H^2 \phi ^2
\zeta[3]}{24 \pi ^4}-\frac{g^3 H^3 \beta ^2
\zeta\left[3,-\frac{\phi }{2 \pi }\right]}{192 \pi ^4}-\frac{g^3
H^3 \beta ^2 \zeta\left[3,\frac{\phi }{2 \pi }\right]}{192 \pi ^4}
+O(\beta^2(gH)^3).
\end{eqnarray}
Here $C_1=-0.01646$, $\gamma\simeq 0.577216$ is Euller's constant,
$\zeta(s)$ is the Riemann Zeta-function, $B_n$ are the Bernoulli
numbers.

We performed the described above lattice calculations and have
obtained the values for the sponta\-neously generated field
strength and the Polyakov loop of corresponding states. Then we
considered the expression (\ref{totHT}) as a function of these
parameters and studied its minimum. We have determined that for
the temperatures $1/\beta=100-500GeV$, the field strengths
$H=10^{22}-10^{24}G$, and $\phi=2.0-3.11$ the one-loop effective
potential (\ref{totHT}) is negative, $V^{(1)}(H,\phi,\beta)<0$.
Hence it follows, either  gauge field or $A_0$ condensate are
spontaneously generated. Moreover, in the minimum of the effective
potential for used values of the parameters the ratio $Im
V^{(1)}/Re V^{(1)}$ is  of order $\sim 10^{-10}$, that is
practically zero for numeric calculations. We conclude that the
common generation of the constant Abelian chromomagnetic field and
the non-trivial Polyakov loop results in the stable magnetized
vacuum state. Of course,  the one-loop potential does not exhibit
the complete effective potential. However, it gives the main
contribution which contains a larger imaginary part. If one adds
the second next-to-leading contribution coming from ring diagrams,
this part is cancelled \cite{SB}. So, the check of stabilization
at one-loop level is important.

\section{Discussion}
We have investigated the  effect on common spontaneous generation
of chromomagnetic field and $A_0$-condensate in
$SU(2)$-gluodynamics at high temperature. The first step  was to
show the possibility of the spontaneous generation of
chromomagnetic field at high temperature \cite{DSlat}. The
obtained results are in a good agreement with that of derived
already in the continuum quantum field theory \cite{SB,SS} and in
the lattice data analysis \cite{AG}. The actual values of the
chrmomomagnetic field strength and the Polyakov loop were obtained
from  the same  MC simulations. They allow to conclude that these
condensates, chromomagnetic field and $A_0$, have to be present in
the deconfinement phase of QCD.

The developed  approach joins the calculation of free energy
functional $F(\phi)$ and the consequent statistical analysis of
its minimum positions for various temperatures and flux values. In
this way the spontaneous creation of condensates is realized in
lattice simulations.

Other important result of the present work is that the
spontaneously magnetized state is stable. This indicates that a
true vacuum at high temperature is formed from  these condensates.

Since the field configuration with  constant fields is gauge
non-invariant, one could consider this configuration as a domain.
A complete structure of whole space can be derived by using the
requirement of gauge invariance for the space including domains
with different orientation. This problem we left for the future.

We would like to conclude that in the deconfinement phase the
condensates have to influence various processes that should be
taken into consideration to have an adequate concept about this
state.

\end{document}